\documentclass[12pt]{article}
\usepackage{scicite}
\usepackage{times}

\newenvironment{sciabstract}{%
	\begin{quote} \bf}
	{\end{quote}}

\usepackage{graphicx}
\usepackage{dcolumn}
\usepackage{bm}
\usepackage{amssymb}
\usepackage{gensymb} 
\usepackage{amsmath}
\usepackage{physics}
\usepackage{float}
\usepackage{xcolor}
\usepackage[english]{babel}
\usepackage{color}
\usepackage{ulem}
\usepackage{soul}

\usepackage{booktabs}
\usepackage{makecell}
\usepackage{hhline}

\usepackage{doi}

\usepackage{siunitx}

\usepackage{colortbl}

\newcolumntype{A}{
 >{$}r<{$}
 @{\extracolsep{0pt}}
 >{${}} l <{$}
 @{\extracolsep{\fill}}
}

\usepackage{caption}

\newcommand{\vkeff}{\vec{k}_{\text{eff}}}
\newcommand{\keff}{k_{\text{eff}}}

\newcommand\stxt[1]{_{\text{#1}}} 

\newcommand{\area}{\vec{\mathcal{A}}}

\newcommand{\deltaK}{\Delta k_{\text{eff}}}

\title{Accurate measurement of the Sagnac effect for matter waves}

\author{
	Romain Gautier$^{1}$, Mohamed Guessoum$^{1}$, Leonid A. Sidorenkov$^{1}$,\\ Quentin Bouton$^{1}$, Arnaud Landragin$^{1}$, Remi Geiger$^{1\ast}$\\
	\\
	\normalsize{$^{1}$LNE-SYRTE, Observatoire de Paris-Universit\'e PSL, CNRS, Sorbonne Universit\'e}\\
	\normalsize{61 avenue de l'Observatoire, 75014 Paris, France.}\\
	\normalsize{$^\ast$To whom correspondence should be addressed; E-mail:  remi.geiger@obspm.fr.}
}

\date{}

\begin{document} 
	\baselineskip24pt
	\maketitle 
	
	\begin{sciabstract}
		A rotating interferometer with paths that enclose a physical area  exhibits a phase shift proportional to this area and to the rotation rate of the frame. 
		Understanding the  origin of this so-called Sagnac effect has played a key role in the establishment of the theory of relativity and has pushed for the development  of precision optical interferometers. 
		The fundamental importance of the Sagnac effect motivated the realization of experiments to test its validity for waves beyond optical, but precision measurements  remained a challenge.
		Here we report the accurate test of the Sagnac effect for matter waves, by using a Cesium-atom interferometer featuring a geometrical area of 11~cm$^2$ and two sensitive axes of measurements. We measure the phase shift induced by the Earth's rotation and find agreement with the theoretical  prediction  at an accuracy level of 25 ppm.
		Beyond the importance for fundamental physics, our work opens  practical applications in seismology and geodesy.
	\end{sciabstract}
	
	\section*{Introduction}
	\label{sec:intro}
	
	The study of the effect of rotations on interferometers dates back to the late nineteenth century and is intimately tied to the development of the theory of relativity. In 1913, Georges Sagnac was the first to report an experimental observation of the shift of the  fringes in an interferometer subject to a constant rotation rate and its interpretation in the framework of an eather theory \cite{Sagnac1913, Post1967,Anderson1994}. Observing the small phase shift induced by Earth rotation motivated Michelson, Gale and Pearson  to build an interferometer of 0.2 km$^2$ area ; in 1925, they reported a measurement of the predicted effect with $3\%$ accuracy \cite{Michelson1925}.
	The advent of the laser boosted the  development of gyroscopes based on the Sagnac effect with the realization of ring-laser gyroscopes \cite{Macek1963} and  later of fiber optical gyroscopes \cite{Vali1976,Lefevre2014}, which are a key component of modern navigation systems.

	The importance of  understanding the fundamental nature of the Sagnac effect for the development of modern physics has motivated the realization of rotating interferometers of increasing precision involving other-than-optical waves.
	
	Observations were subsequently made with various systems, starting with superconducting electrons  \cite{Zimmerman1965} as one of the first demonstration of a macroscopic matter-wave coherence in superconductors. It was followed by measurements with neutral particles: first with neutrons \cite{Werner1979} and then with thermal atoms \cite{Riehle1991}, where the Sagnac effect was found to be in good agreement with theory. A measurement with electron jet \cite{Hasselbach1993} has extended its validity towards matter-waves of free charged particles. Study of the Sagnac effect in superfluid quantum liquids (Helium 4 Ref\cite{Schwab1997} and Helium 3 Ref\cite{Simmonds2001}) and gases (BEC \cite{Wright2013}) has illustrated its universality. 
		These proof-of-principle experiments served to underline the relativistic nature of the Sagnac effect.
	The first precision measurement was done in 1997 , with a reported accuracy of 1\% for a thermal matter wave interferometer\cite{Lenef1997}. 
		Development of cold atom experiments allowed for measurements of increasing precision \cite{Gauguet2009, Stockton2011} up to 0.05\% preceding this work.

	According to the Sagnac effect, the phase shift in an interferometer of oriented area $\vec{\mathcal{A}}$ and subject to a constant rotation rate $\vec{\Omega}$ can be expressed as
	\begin{equation}
	\Phi_{\Omega} = \frac{4\pi E}{h c^2} \area\cdot \vec{\Omega},
	\label{eq:sagnac_phase}
	\end{equation}
	where $E$ is the total energy of the interfering particle and $h$ the Planck's constant ($E=h\nu$ for photons, $E\simeq m c^2$ for slow massive particles).
	Precisely testing the validity of this equation requires an accurate knowledge of the  interferometer geometry (i.e. of the area vector $\area$) and of the  rotation rate ($\vec{\Omega}$).  Exploiting Earth rotation, which is known with high accuracy, meets the latter requirement. However, precisely controlling the geometry of a matter-wave interferometer of large area (i.e. of high sensitivity) remains a challenge; for example, the accuracy of superfluid Helium interferometers has been barely assessed \cite{Sato2014}, while neutron interferometers could test Eq.~\ref{eq:sagnac_phase} at best with $0.4\%$ accuracy \cite{Werner1979}.

	Cold-atom interferometers feature a  high degree of accuracy  owing to the good knowledge of the light-matter interaction process exploited to realize the interferometer building blocks, which offers the possibility to quantify the interferometer scale factor  using frequency measurements \cite{Barrett2014,Geiger2020}.
	Here, we use a two-axis cold-Cesium atom interferometer with a macroscopic area $\mathcal{A}\simeq 11$~cm$^2$ (in each direction) 
	rotated by the Earth. 
	Our measurements confirm the prediction of Eq.~\eqref{eq:sagnac_phase} with an accuracy of 25 ppm, which represents an improvement of more than 20 compared to previous experiments \cite{Gauguet2009,Stockton2011}, and allows us to place a constraint  on G\"{o}del's model of a rotating Universe  \cite{Delgado2002}. 
	Moreover, the ability to accurately determine the scale factor of our gyroscope combined with its relative compactness and control of its area orientation (compared to giant ring laser gyroscopes \cite{Gebauer2020}),  opens practical applications in seismology and geodesy. 
	The configuration of our instrument has advantages over other geometries of cold atom gyroscopes. It's single source folded interferometer rejects accelerations while preventing systematic errors due to trajectories mismatch in twin atom-source sensors \cite{Gauguet2009, Berg2015, Rakholia2014, Yao2021}. 
		Our sensor offers significant sensitivity gain compared to that of compact atomic gyroscopes, \cite{Chen2019, Alzar2019, Wu2017}, allowing us to test the Sagnac effect due to Earth rotation with the accuracy level reported in this work.
	\section*{Results}
	\label{sec:principle}

	\begin{figure}[hbtp]
		\centering
		\includegraphics[width=1\columnwidth]{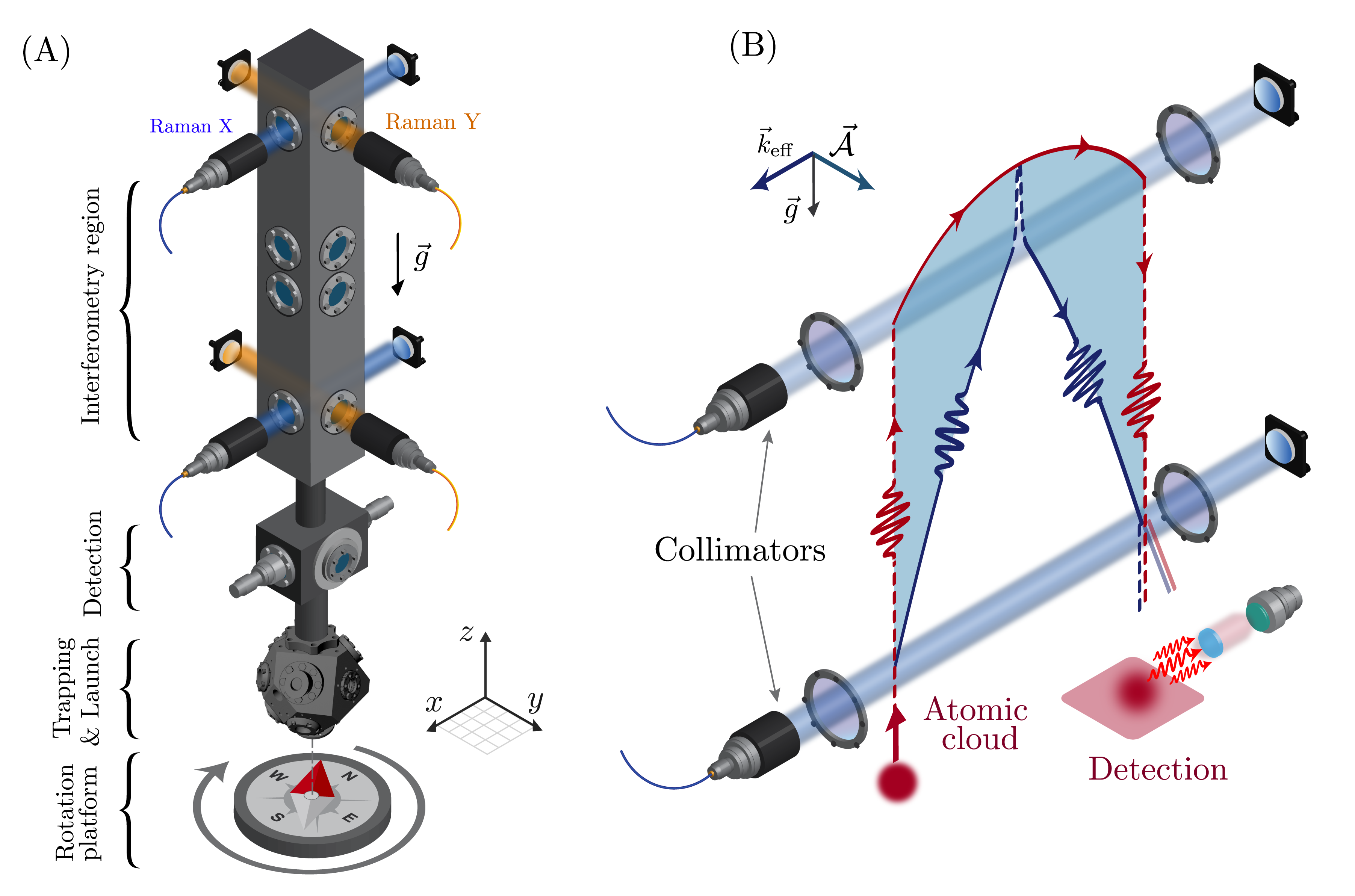}
		\caption{\textbf{Principle of the experiment}. (A) Schematic of  the sensor head. In the lower part of the vacuum chamber the Cesium atoms are laser-cooled and trapped  in a Magneto-Optical Trap (MOT),  and then launched vertically in the hyperfine state $F=4$. Subsequently, the atoms enter the interferometer  where a sequence of four Raman transitions is driven by retro-reflected laser beams at the top and the bottom of the upper part of the vacuum chamber. The interferometer can be operated either in X (blue beams) or in Y (orange beams) direction. 
			At the output of the interferometer, the probability for an atom to occupy one of the two internal states $F=3$ and $F=4$ is measured by fluorescence detection. 
			The experiment is placed on a rotation stage   which allows us  to vary the projection of the oriented interferometer area on the Earth rotation vector. 
			(B) Schematic of the wavepacket propagation  in the interferometer (here in X direction, not to scale). The red and blue lines show the two distinct paths of the splitted matter-waves enclosing a physical area, underlined by the cyan color. Dashed and plain lines encode the two internal states of the atom.}
		\label{fig:experiment}
	\end{figure}

	The core of our experiment and its principle are illustrated in Fig. \ref{fig:experiment} and has been described in previous works ~\cite{Dutta2016,Savoie2018} and summarized in the Materials and Methods. The interferometric sequence comprises four Raman pulses of Rabi area $\pi/2,\pi,\pi,\pi/2$ occurring at times $t\simeq (0, T/2, 3T/2, 2T)$, with $T\simeq 400$~ms.  
	
	The two-photon Raman transition transfers a momentum $\hbar \vkeff$ to the deflected atom, which, together with the action of gravity acceleration $\vec{g}$ along the path, results in an interferometer area (for perfectly parallel Raman beams):
	\begin{equation}
	\area=\frac{T^3}{4}\frac{\hbar}{m}\vkeff \times \vec{g}.
	\end{equation}
	With the total energy of the interfering atom given by $E\simeq mc^2$ (valid for  atoms moving much slower than light), the Sagnac phase shift becomes
	\begin{equation}
	\Phi_{\Omega}=\frac{T^3}{2} (\vkeff \times \vec{g}) \cdot \vec{\Omega}.
	\label{eq:sagnac_keff}
	\end{equation}
	
	The Raman beams are set to an angle $\theta_0$ with respect to the horizontal plane (perpendicular to $\vec{g}$) in order to lift the degeneracy associated with the two possible directions of momentum transfer and thereby choose the direction of atom diffraction $[\pm \vkeff]$. The vector product is then expressed as $\vkeff \times \vec{g}=\keff g \cos(\theta_0) \vec{n}$, where $\vec{n}$ is a unit vector in the direction of interferometric area ($\area/|\area|$) lying in the horizontal plane.

	The experiment is placed on a rotation stage, which allows us to change  the angle between $\vec{n}$ and the angular velocity of the Earth $\vec{\Omega}$ pointing from South to North. The rotation angle $\Theta$ can be varied within $2\pi$ with $\mu$rad accuracy, thus permitting a precision measurement that is not limited by uncertainty in positioning of the North.

	The Sagnac phase shift can therefore be explicitly written in terms of the control parameters as 
	\begin{equation}
	\Phi_{\Omega}(\Theta)=\frac{T^3}{2} \keff g \cos(\theta_0) \times  \cos(\psi) \Omega_E \times \cos(\Theta - \Theta_N),
	\label{eq:sagnac_expanded}
	\end{equation}
	where $\Omega_E$ is the modulus of the Earth rotation vector, $\psi$ is the astronomic latitude at the position where the experiment is performed on the site of Paris Observatory, and $\Theta_N$ is the angle of the rotation stage corresponding to geographical North.

	We realize two independent measurements with Raman beams oriented in the X and Y directions, i.e.  with interferometer areas perpendicular to each other (Fig.~\ref{fig:experiment}). The two interferometers operate on the same physical principle (Eq. \ref{eq:sagnac_keff}) but with different scale factor and bias term thus increasing our confidence in the final result.

	The phase shift measured at the output of the atom interferometer, $\Phi$, is dominated by the rotation-induced Sagnac term of interest (of the order of 200~rad) and contains other bias terms on the order of few tens of mrad detailed in Materials and Methods.

	\begin{figure}[htbp]
		\centering
		\includegraphics[width=1\columnwidth]{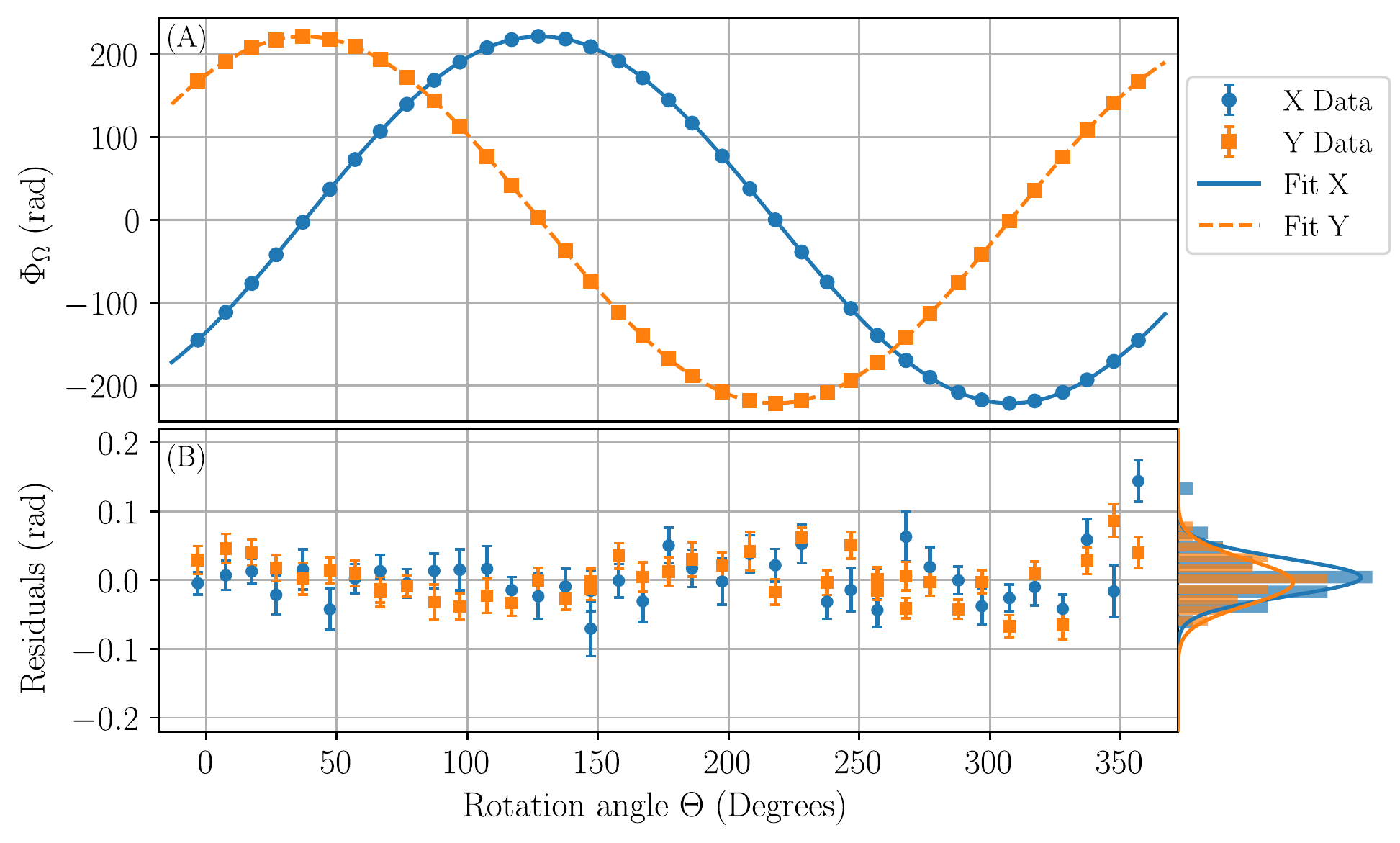}
		\caption{\textbf{Measurement of the Sagnac phase shift with the two-axis atom interferometer}. (A) Phase shift acquired for the X (blue dots) and Y (orange squares) directions as a function of the rotation angle $\Theta$. Each point is a mean of typically 1500 realizations, with statistical error smaller than the symbol size. Lines are the least-squares fits with Eq.~(\ref{eq:sinus_fit}). (B) Difference between the data and the fits for X (blue dots) and Y (orange squares). Statistical error of each point is on the order of 30 mrad and the histogram of the residuals (projected on the right side) has a standard deviation of 40~mrad.}
		\label{fig:sinus}
	\end{figure}

	Fig.~\ref{fig:sinus} shows a typical measurement of the phase shift of the atom interferometer for both directions, acquired during one week \textit{in April 2021}. Despite the interferometer measuring a phase shift modulo $2\pi$ in a given orientation, the complete 360$^{\circ}$-variation of the rotation angle allows us to unambiguously "unfold" the full $\sim200$~rad dephasing~(see Supplementary Materials). The data are fitted with
	\begin{equation}
	\Phi_{\Omega}^{x,y}(\Theta) = \Phi_0^{x,y}\cos(\Theta-\Theta^{x,y}_N) + B^{x,y},
	\label{eq:sinus_fit}
	\end{equation}
	where $\Phi_0^{x,y}$, $\Theta^{x,y}_N$ and $B^{x,y}$ are free parameters (3 for each direction).
	We extract $\Phi_0^x = 221.572(9)$ rad and $\Phi_0^y = 221.545(9)$ rad with fit residuals characterized by histograms with Gaussian width of about 40~mrad, comparable with the error bar of the individual points. Additional deviation in fit residuals can be explained by slow drifts of the bias during the measurement (see Supplementary Materials).  The extracted relative angular mismatch of X and Y directions from being perfectly orthogonal is found to be 0.7(1) mrad, compatible with mechanical tolerance on the orthogonality of the sides of the vacuum chamber.

	We now estimate the gyroscope scale factor for both directions (X and Y), i.e. evaluate the parameters entering Eq.~\eqref{eq:sagnac_expanded}.
	As we will show, all the parameters can be determined solely by frequency (or time) measurements, i.e. with high accuracy.

	We measured the angle $\theta_0^{x,y}$ with the four-pulse interferometer by exploiting its residual sensitivity to DC acceleration (see Supplementary Materials), and obtained $\theta_{0}^x$ = 4.0750(5)$^{\circ}$  and $\theta_{0}^y$ = 4.1251(3)$^{\circ}$. 
	The interrogation time is  derived from the clock of the experimental control system that is referenced to a highly stable and reproducible frequency standard. To check for small possible systematic deviations, we measure  the time interval between the Raman pulses with a high speed oscilloscope and find a value $T=400.0020(1)$~ms, with error bar limited by available temporal resolution.
	The local gravity acceleration value $g$ has been previously measured in the laboratory using a transportable cold atom gravimeter. Since the value of $g$ was affected by tides during the present measurement campaign, we take the maximal annual tide-induced variation of  $3 \times 10^{-6}$~m.s$^{-2}$ as an upper bound for the uncertainty on the value of $g$.

	At the level of accuracy of typically 50~ppm (as demonstrated by presented single data set), we must account for the fact that the modulus of the wave-vectors for the bottom ($k_{\mathrm{eff}}^{(B)}$) and top ($k_{\mathrm{eff}}^{(T)}$) Raman beams might differ by $\Delta k_{\mathrm{eff}} \equiv k_{\mathrm{eff}}^{(B)} - k_{\mathrm{eff}}^{(T)} $, which introduces a correction to Eq.~\eqref{eq:sagnac_expanded} at first order in   $\epsilon = \Delta k_{\mathrm{eff}} / k_{\mathrm{eff}}$  \cite{Sidorenkov2020}.
	
	We measure the values of $\epsilon^{x,y}$ via an interferometric measurement as explained in the Supplementary Materials, leading to  $\epsilon^x$ = $0.7(9) \times 10^{-6}$  and $\epsilon^y =6.0(9) \times 10^{-6}$.

	Finally, we evaluate the astronomical latitude $\psi$ -the angle between the local vertical (i.e. the vector perpendicular to the geoid) and the equatorial plane. It differs from the geographic latitude by the vertical North deflection, which can reach several arc-seconds in  regions where the geoid deviates significantly from the ellipsoid of reference (e.g. close to mountains). Our experiment is positioned in a room of Paris Observatory, at the geographic latitude of $48.83561(2)^{\circ}$. The vertical deflection was estimated to $+0.95(4)^{''}$, yielding $\psi = 48.83587(3)^{\circ}$.
	
	Second order ($\propto \Omega^2$), recoil ($\propto \hbar^2 \keff^2/(2m)$) and other residual terms appear in the expression of the Sagnac phase shift when the full calculation is developed (see Supplementary Materials). 
	These contributions would correspond to a relative correction of few~$10^{-7}$ to the estimation of the scale factor, which is two orders of magnitude below the accuracy of our measurement and have thus been neglected in this study.

	Table~\ref{table:expected} summarizes the measurements of the parameters for both directions. Based on these measurements, we estimate the theoretical values for the Sagnac phase shift as $\Phi\stxt{theo}^x=221.5702(3)$~rad and $\Phi\stxt{theo}^y=221.5574(2)$~rad.

	\begin{table}[htb]
		\centering
		\renewcommand*{\arraystretch}{1.3}
		\begin{tabular}{cc|c|c}
			\toprule
			\multicolumn{1}{c}{Parameter} 		& X & Y  & \makecell{Relative \\ uncertainty (ppm)}\\
			\midrule
			\multicolumn{1}{c}{$k_{\mathrm{eff}}$ }		& 	\multicolumn{2}{c}{$14743247.08(4)~\text{m}^{-1}$} 	&  $0.003$\\
			\multicolumn{1}{c}{$T$}			& \multicolumn{2}{c}{$400.0020(1)$~ms}  & $0.75$\\
			$g$				&	\multicolumn{2}{c}{$9.809279(3)~\text{m.s}^{-2}$}	& $0.3$\\
			\rule{1pt}{0ex}
			$\theta_0$ 		& \multicolumn{1}{c|}{$4.0750(5)^{\circ}$} & \multicolumn{1}{c}{$4.1251(3)^{\circ}$} & $ 0.6~|~0.4$\\
			$\epsilon$ 	& \multicolumn{1}{c|}{$-1.7(1) \times 10^{-6}$}&  \multicolumn{1}{c}{$-9.3(2) \times 10^{-6}$}&  \multicolumn{1}{c}{$0.07 ~|~ 0.13$} \\
			$\psi$	& 	\multicolumn{2}{c}{$48.83587(3)^{\circ}$ }	&  $0.6$\\
			$\Omega_E$		& 	\multicolumn{2}{c}{$7.2921150(1) \times 10^{-5}~\text{rad.s}^{-1}$} 	& $0.01$\\		
			\midrule
			\rule{0pt}{4ex}  
			\makecell{\textbf{Theoretical} \\ \textbf{Sagnac Phase}} 	& \multicolumn{1}{c|}{221.5702(3) rad} & 221.5574(2) rad & $1.2~|~1.1$\\
			\bottomrule
		\end{tabular}
		\caption{\textbf{Error budget for the determination of theoretical Sagnac phase shift.} The table lists the parameters entering the scale factor of the cold-atom gyroscope. The right column is the uncertainty on the scale factor resulting from error propagation on each parameters.}
		\label{table:expected}
	\end{table}

\begin{figure}[htbp]
		\centering
		\includegraphics[width=0.8\columnwidth]{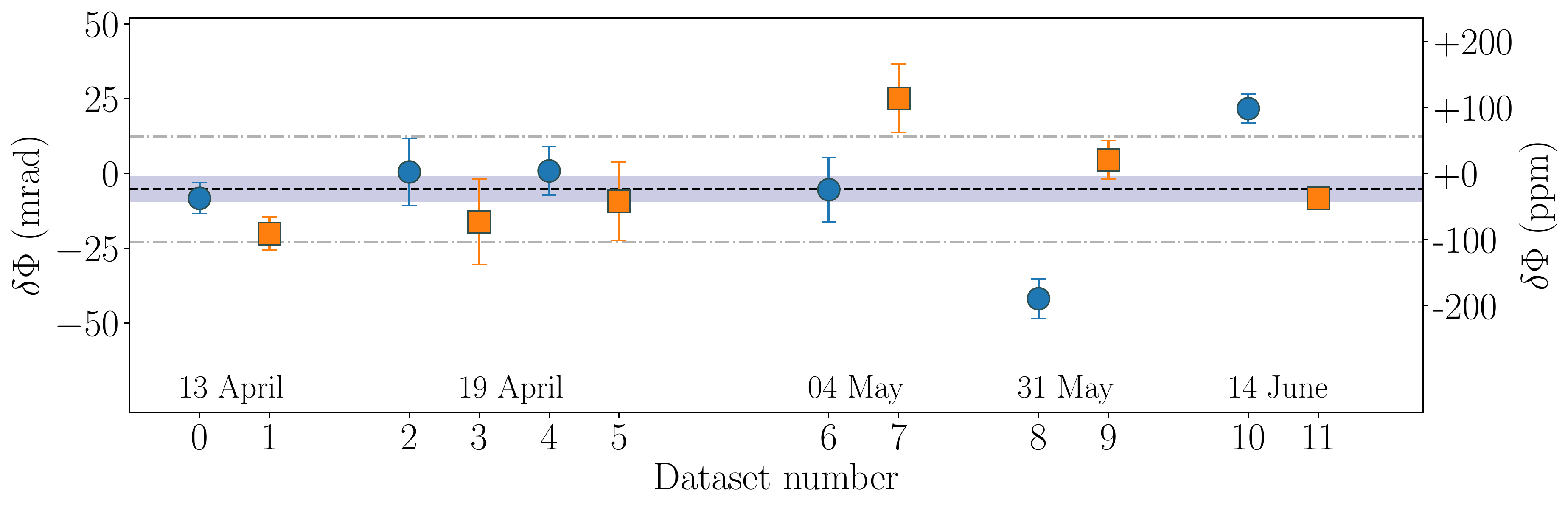}
		\caption{\textbf{Comparison between experiment and theory.} The data points represent the difference between the  measured gyroscope phase shift and the theoretical Sagnac phase shift. 	Blue points (orange squares) are the data for the X (Y) axis. The dashed line represents the mean value of all the data (X and Y) and the shaded region is the standard error on the mean. Point-dashed lines indicates the standard deviation on the set of measurement. The week  of data set acquisition (during year 2021) is indicated at the bottom along with the dataset number.}
		\label{fig:all_measurements}
	\end{figure}

	To reinforce the overall confidence in our measurements, we acquired six full-turn data from April to June 2021 and with different experimental parameters (e.g. variation of interrogation time $T$). We applied identical data treatment and fitting procedures (as described above) to all data sets and extracted corresponding fit parameters $\Phi_0^{x,y}$ (see supplementary materials for the raw data and full fit results). In Fig.~\ref{fig:all_measurements} we present the differences between the measured values and the corresponding theoretical expectations, $\delta\Phi^{x,y} = \Phi_0^{x,y}-\Phi\stxt{theo}^{x,y}$.
	
	The results have an overall good agreement, with a mean value close to zero (horizontal dashed line) and a standard error on the mean covering significant part of the data (gray-shaded region). The dispersion between the measurements is not fully captured by the uncertainties in fitted values of $\Phi_0^{x,y}$ of the corresponding data sets. We show in Supplementary Materials with additional simulations that the deviations are consistent with a residual shift of the bias during the week-long measurements necessary to rotate the experiment.
	In conclusion, the data are consistent with the Sagnac phase shift prediction within an uncertainty of 25 ppm, dominated by the statistical uncertainty.
	
	\section*{Discussion}
	\label{sec:discussion}
	
	The fundamental interest in our cold-atom interferometer relies on its ability to measure several components of the local angular velocity and to explore rotational signals along different directions, in contrast to  large ring laser gyroscope infrastructures where the gyroscope axes are fixed. This allows to search for smaller signals beyond Earth rotation: in addition to local angular velocities induced by geological origins, one can constrain astrophysical rotations (orbit in the Solar System, rotation in the Galaxy) or even rotations related to the fundamental structure of the Universe. As shown in Ref.~\cite{Delgado2002}, G\"{o}del's model of the Universe predicts a global rotation rate inducing a Sagnac phase shift.  The mass density of the Universe inferred from the 2018 Planck mission data \cite{Planck2020} corresponds to a rotation rate of order $10^{-19}$~rad.s$^{-1}$, far beyond the accuracy of current gyroscopes. However, the accuracy of our experiment gives for the first time an upper limit on the G\"{o}del's rotation obtained for matter-waves (instead of photons in the case of the Planck mission), in a local measurement. From another point of view, the ability to detect signals at different frequencies in the experiment is a powerful tool to explore violation of Lorentz invariance. Following Ref.~\cite{Moseley2019}, our experiment might put constraints on parameters of alternative theories such as the Standard Model Extension.

	Additionally, the precise knowledge of the scale factor of our gyroscope together with that of the Earth rotation rate allows us to perform measurements of the vertical deflection at the level of few arcseconds. This provides a measure of the local gravity direction which depends on local mass anomalies.  Accurate knowledge of the vertical deflection (which can amount to angles of few arcsec in flat areas and up to $50^{\prime \prime}$ in mountainous terrain) is widely used in geodesy and for geophysical purposes. A high accuracy gyroscope such as ours allows to measure at least the North-South component of the vertical deflection in the zones where astronomical determination is impossible, e.g. for geodesy and geographical positioning in underground facilities.

	Finally, our work paves the way towards applications in rotational seismology, a field that studies rotational motions induced by earthquakes, explosions, and ambient vibrations \cite{Lee2013}, of interest for the understanding of the  underground structure \cite{Bernauer2009} or   seismic hazard assessment in civil engineering \cite{Basu2015}. Theoretical studies have shown the benefit  of using precision rotational sensing to improve the characterization of earthquake sources \cite{Donner2016} and their localization \cite{Li2017}- the information of prime importance for seismic alert systems. 
	
	Accurate assessment of ground rotational signals is also of prime importance in the development of ground-based gravitational wave detectors, which is expected to be limited at low frequencies (below 1 Hz) by Newtonian noise \cite{Harms2020, Canuel2020} and rotational ground motion \cite{Ross2020}.
		Mesuring and compensating these effects requires highly sensitive and accurate rotational sensors on the level of performances of our instrument\cite{Lantz2009}.
	
	Such measurements at geophysical sites of interest require transportable gyroscopes with scale factors that are stable over weeks and are known with high accuracy (better than 100 ppm). While fiber-optic gyroscopes \cite{Bernauer2018} have been particularly developed and deployed for rotational seismology applications, reaching such stability and accuracy levels is challenging. Our cold-atom sensor could lead to a transportable laboratory instrument \cite{Farah2014} or even to an industrial product with increased robustness against environmental instabilities (temperature, vibrations etc.), as achieved for cold-atom gravimeters \cite{Menoret2018}. A specific effort should address the control of the bias drift of our gyroscope due to the fluctuations of atomic trajectory coupled to relative mirror misalignment \cite{Altorio2020}, that appeared as a limiting factor in the present work (see Supplementary Materials). Provided such proper engineering, the level of accuracy reported by our work opens this new field of applications, with major scientific and societal impacts.

	\section*{Materials and Methods}
	\subsection*{Preparation and detection of the atoms}
	Cesium atoms laser-cooled to a temperature of 1.8~$\mu K$ are launched vertically in a fountain configuration at a velocity of 5~m.s$^{-1}$. After a quantum state selection in the least sensitive magnetic sub-level $\ket{m_F=0}$, the atoms enter the light-pulse  interferometer, where a sequence of stimulated two-photon Raman transitions split, deflect and recombine  the atomic de Broglie waves.
	
	At the output of the interferometer, the phase difference between the two paths is inferred by measuring the internal state (entangled with the external state) populations of the atoms via fluorescence detection. We operate the interferometer in joint mode \cite{Dutta2016} such that the time of a full cycle equals the total interrogation time $2T\simeq 800$~ms. 
	
	\subsection*{Alignement of the interferometer}
	We use a dedicated alignment protocol~\cite{Altorio2020} allowing to set the atomic launch velocity parallel to vertical (local $\vec{g}$) with an accuracy of typically 200~$\mu$rad. Once set, this alignment is preserved upon variation of the rotation angle during the full-turn acquisition by actively stabilizing the sensor's tilt at the nrad-level during the acquisition~(see Supplementary Materials).
	
	\subsection*{Phase shift of the interferometer}
	We write the total phase shift at the output of the interferometer as $\Phi=\Phi_\Omega + \Phi_1 + \Phi_2$, where $\Phi_1$ and $\Phi_2$ respectively encode $\vkeff$-independent and $\vkeff$-dependent bias phase shifts.
	The contribution of $\Phi_1$, mostly one-photon light shift is maintained below 10 mrad by alternating measurements every  cycle  between $+\keff$ and $-\keff$ and computing the half-difference between the data (see Supplementary Materials).
	
	The most important terms contributing to $\Phi_2$ are: \textit{(i)} a DC acceleration-induced phase shift \cite{Stockton2011,Sidorenkov2020}; \textit{(ii)} a phase shift associated with the imperfect alignment  of the bottom and top mirrors retro-reflecting the Raman beams (see Fig.~\ref{fig:experiment}) coupled to imperfect launching of the atoms along gravity \cite{Altorio2020}. We recall in Supplementary Materials the origin of these phase shifts and explain the methods employed to mitigate their contribution all along the measurements,  for both X and Y directions.
	
	

	\clearpage


	\section*{Acknowledgments}
	We thank Matteo Altorio for his contribution at the early stage of this work, Marie-Christine Angonin and Christophe Le Poncin-Lafitte for stimulating discussions and Franck Pereira Dos Santos for careful reading of the manuscript. We thank as well J. Gazeaux of the Service de G\'eod\'esie et de M\'etrologie department of the IGN for giving us the precise value of the North deviation at the location of the Paris Observatory.

	\noindent\textbf{Funding:} We acknowledge the financial support from Agence Nationale pour la Recherche (project PIMAI, ANR-18-CE47-0002-01) and Centre National d'Etudes Spatiales (CNES). L.A.S. was funded by Conseil Scientifique de l'Observatoire de Paris (PSL fellowship in astrophysics at Paris Observatory),  R. Gautier by the EDPIF doctoral school and M.G. by  SIRTEQ. 
	 	
	\noindent\textbf{Author contributions:} R.Gautier, M.G., L.A.S. performed the experiments.
	R.Gautier, M.G., L.A.S. and Q.B analyzed the data and did the calculations.
	R.Geiger and A.L. conceived the experiment and supervised the research.
	R. Gautier and R. Geiger wrote the manuscript, with inputs from all the authors.
	All authors discussed the manuscript.
	
	\noindent\textbf{Competing Interests:} All other authors declare they have no competing interests.
		
	\noindent\textbf{Data and materials availability:} All data needed to evaluate the conclusions in the paper are present in the paper and/or the Supplementary Materials. Data and code are available at  \url{https://doi.org/10.5281/zenodo.6372385}.

	\section*{Supplementary materials}
	Supplementary Text, Sections S1 to S4\\
	Figs. S1 to S8\\
	Tables T1 to T3\\
	5 References.

	\clearpage


%

\newpage

\begin{center}
\Large{\textbf{Supplementary Materials}}
\end{center}

\renewcommand{\thefigure}{S\arabic{figure}}
\renewcommand{\thetable}{T\arabic{table}}
\setcounter{figure}{0} 
\setcounter{table}{0} 

\section*{Section S1: Experiment details and analysis of data}
\subsection*{Section S1.1: Single measurement}

\subsubsection*{Extraction of $\Phi_\Omega$}

Each phase measurement is extracted by alternating between two side of the interferometric fringe, in a way to be locked at the center and always stay in the linear range, subsequently getting rid of contrast and offset fluctuations.

Contributions from non-inertial phase-shifts $\Phi_1$ are reduced by a factor 100 by alternating interleaved measurement of $+\keff$ and $-\keff$ momentum transfer.

Inertial phases shifts present in $ \Phi_2$ are dependent on the sign of $\keff$ and are the following:

First, a DC acceleration term originating from the asymmetry introduced on the 4 pulses geometry to prevents parasitic interferometers from recombining. The two middle $\pi$ pulses are shifted by a quantity $\Delta T_a$ in the same direction (thus asymmetrically with respect to the apogee), producing a phase shift:

\begin{equation*}
	\Phi_{DC} = 2T\Delta T_a k_{\mathrm{eff}} (g \sin  \theta_0 - \alpha)
\end{equation*}
where $\Delta T_a$ is the asymmetric timeshift introduced, $g \sin  \theta_0$ is the projection of the gravity on the Raman beams, and $\alpha$ represents the rate of the frequency ramp applied on the Raman lasers to compensate for the variation in Doppler effect during the intereferometer.
The ramp is first adjusted in a way to cancel the effect to a large extent, from 200 rad to hundreds of mrad. 

To reject it even more, we alternate measurements of $+\Delta T_a$ and $-\Delta T_a$ to remove any residual sensitivity to DC acceleration.

The final phase is extracted from both the alternation of the sign of $\vkeff$ and  $\Delta T_a$ as follow :

\begin{equation*}
	\Phi_{\Omega} = \frac{1}{2} \left[\left(\frac{\phi^{+k}_{+\Delta T_a} - \phi^{-k}_{+\Delta T_a}}{2}\right) + \left(\frac{\phi^{+k}_{-\Delta T_a} - \phi^{-k}_{-\Delta T_a}}{2}\right) \right]
\end{equation*}

\begin{figure}[htbp]
	\centering
	\includegraphics[width=0.8\columnwidth]{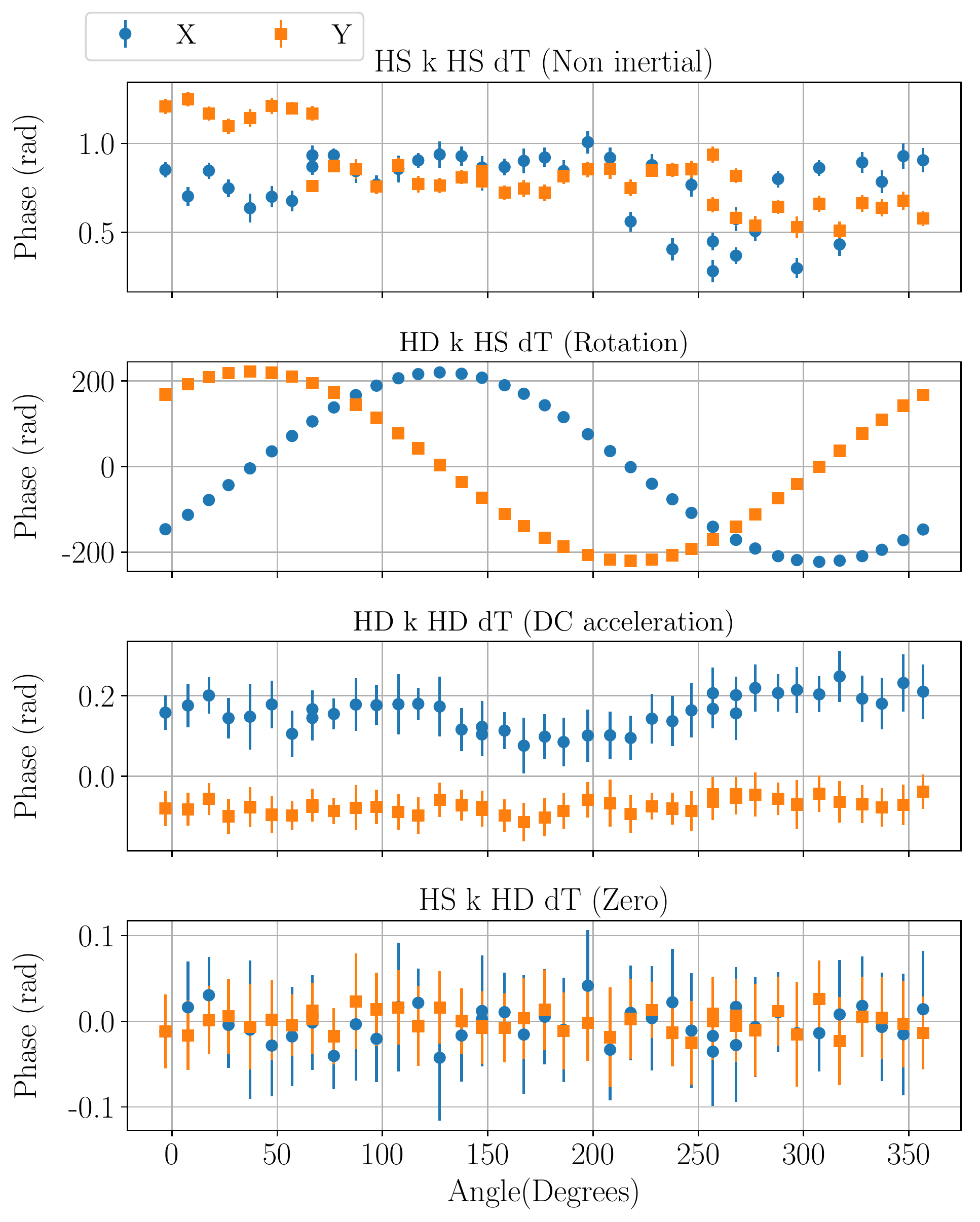}
	\caption{\textbf{Separated signals for a single dataset for the two axis.} The dataset correspond to the one presented on Figure 2 from the main text. From the different combinations of signals, we can extract multiple phase information. From the half sum of $\keff$ and the half difference of $\Delta T_a$ comes the non inertial terms. In the half difference of $\keff$ and the half sum of $\Delta T_a$ comes the Sagnac phase shift. In the half difference of $\keff$ and the half difference of $\Delta T_a$ comes the DC acceleration. From the half sum of $\keff$ and the half difference of $\Delta T_a$, no terms should be left, leaving a constant zero phase shift.}
	\label{fig:separatedSignals}
\end{figure}

The non inertial terms will appears in the half sum of the $+\keff$ and $-\keff$ signals and the terms from the DC acceleration appears in the half difference of $+\Delta T_a$ and $-\Delta T_a$ signals for opposite $\keff$ signals. Examples of the useful and rejected signals are presented in figure \ref{fig:separatedSignals}.

\subsubsection*{Alignement of the Raman beams}
Another inertial signal which need to be reduced is due to the coupling between the alignment of the two Raman mirrors and the trajectory of the atoms (it appears if atoms are not launched perfectly along the gravity acceleration vector).

By repeating the procedure described in (31) we are able to align the two mirrors at the level of $3 \mu \text{rad}$, and adjust the vertical and transverse velocity with an accuracy of $1 \text{mm.s}^{-1}$, thus limiting this systematic effect at the beginning of the measurements at the level of 12 mrad.

\subsubsection*{Vibrations}
To prevent the intereferometric signal from being impaired by the vibrations, the noise is reduced by hybridizing our atomic interferometer with two classical sensors (30). 
The two seismometers (3 Axis Trillium Compact 100 Hz, 120s bandwidth) are rigidly fixed to the Raman retro-reflection mirrors of the X axis and the vacuum chamber, and are used to record the vibrations during the interferometer. The three axis sensors allow us to extract vibration both for the X and Y measurements.

The signal measured by the two seismometers are combined with the transfer function of the experiment to estimate the phase shift due to vibrations (30, 47), which is fed-back to the Raman lasers in real time to be compensated. This reduce the vibration noise contribution to around 300 mrad phase noise per shot, equivalent to a rotation stability of $7\times 10^{-8} \text{ rad.s}^{-1}/\sqrt{\tau}$.

\subsection*{Section S1.2: Control of the rotation angle}

The whole experiment is mounted on a floating platform (Minus-K) which is itself mounted on a rotation stage (ALAR-250LP by Aerotech), and can be rotated by 360$^{\circ}$ with an accuracy of 10 $\mu \mathrm{rad}$. The rotation platform is not automatized and it's electronics used only for a readout of the relative angle of the experiment $\Theta$.

When the experiment is rotated, it can happen that the angle between the rotation stage and the vibration isolation platform differs by a non negligible amount due to some twist or cables pulling on the experiment (it can reach up to $\pm 5 \mathrm{mrad}$) .

\begin{figure}[htbp]	
	\centering
	\includegraphics[width=0.7\textwidth]{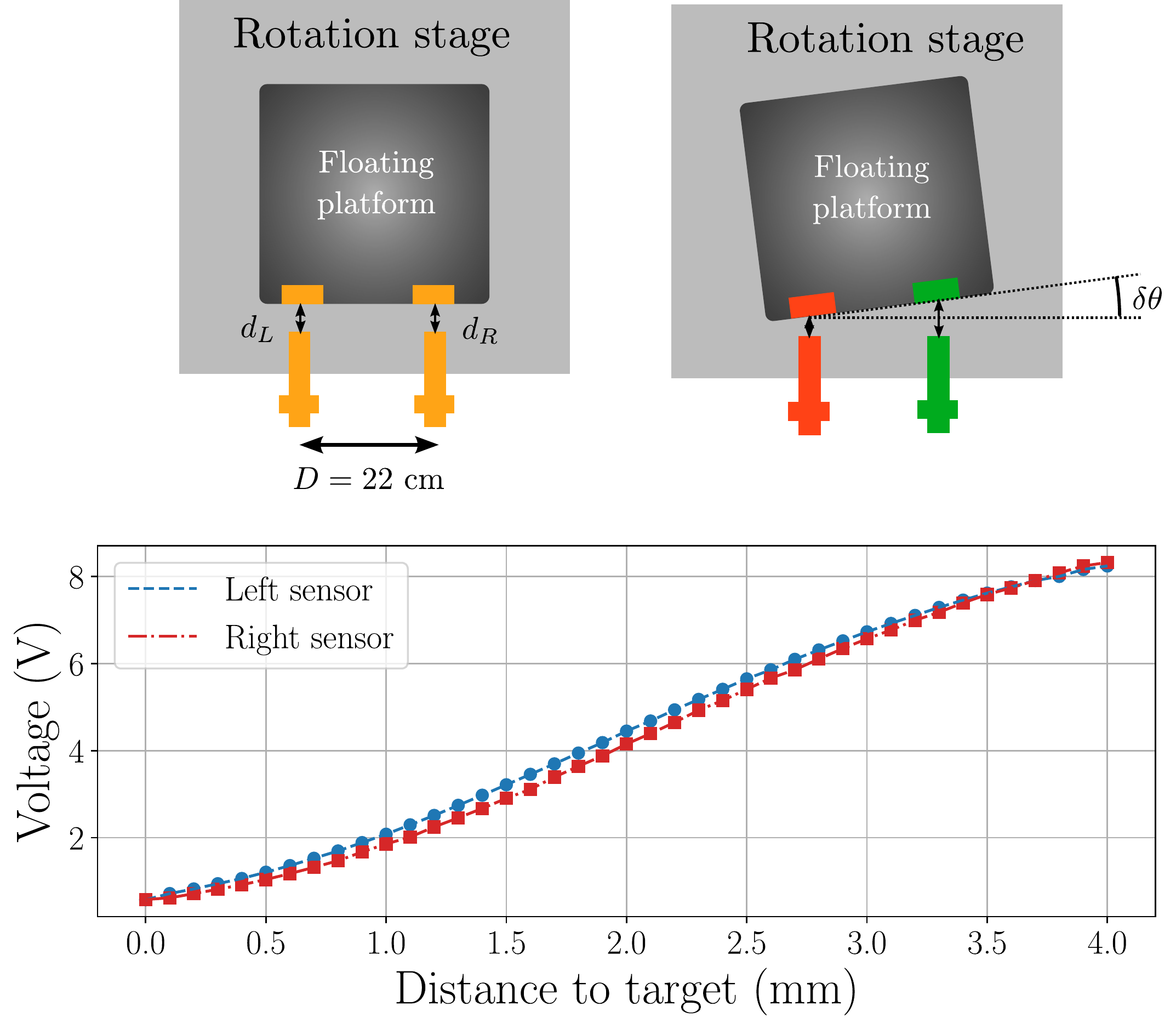}	
	\caption{\textbf{Schematics of the rotation stage, the floating platform, and the two position sensors used to estimate the relative rotation $\delta \theta$ between the stage and the platform.} Bottom figure presents the characterization of the two proximity sensors, which have a slightly different response. From the voltage measured by the two sensors, we can deduce the angle between the rotation stage and the floating platform with $10~\mu \text{rad}$ accuracy.}
	\label{fig:positionSensors}
\end{figure}

To measure the angle between the two platforms, we installed two inductive proximity sensors (Contrinex DW-AS-509-M12-390) separated by a baseline of 22.0(1) cm.
The two sensors are fixed on the rotation stage, while the two steel target blocks are fixed in front, on the floating platform. By reading the voltage of the two sensors and using the calibration curve showed in Fig \ref{fig:positionSensors}, we can deduce the respective distance between the sensors and the platform, allowing us to measure the angle by which the experiment effectively rotated, with $10~\mu \text{rad}$ accuracy.

\subsection*{Section S1.3: Details on one data-set}

For each orientation of the experiment, we integrate the measurement over 20 min, once for the X axis, and once for the Y axis. Error bars on each points corresponds to the standard deviation on the measurement.

Once we have a measurement for the two axis for a given orientation of the experiment, we change the orientation by steps of 10 or 20 degrees, or smaller depending on the dataset.

In order to prevent, as much as possible, the impact of a potential drift of the gyroscope bias on the quality of a dataset, the experiment was not rotated in one single turn from 0$^{\circ}$ to 360$^{\circ}$. Instead, for example, we rotated the experiment in one way from 0$^{\circ}$ to 360$^{\circ}$ by steps of 20$^{\circ}$, and back from 350$^{\circ}$ to 10$^{\circ}$ by steps of 20$^{\circ}$ (i.e. interleaved with respect to the first way), leading effectively to an angular increment of 10$^{\circ}$. Such a procedure reduces the impact of correlated noises on the measurement, while keeping the number of effective rotation of the experiment lower than if the rotation angles were randomly selected.

As the amplitude of variation of the measured phase is much greater than $2 \pi$, at first, only the top of the sinusoid is used to fit the data where we can stay in a $2\pi$ range.
The following data-points are adjusted with an even number of $\pi$ automatically added to match the first fit.

\subsection*{Section S1.4: Details on the full datasets}

Six distinct datasets have been taken both for X and Y axes.
Each full measurement for the two axes takes 1 week, with around 10 rotation of the apparatus per day.
The extracted amplitude, phase and offset are presented on the table below.

For the datasets 2, 3, 4 and 5, instead of taking one measurement for X and one measurement for Y, we took alternatively a measurement for X with 2T = 800ms, a measurement for X with 2T = 801 ms and the same for Y. To be able to complete the measurements in one week, we did only half the rotation angles for each datasets.

\begin{table}[htbp]
	\centering
	\renewcommand*{\arraystretch}{1.2}
	\begin{tabular}{ccccclllc}
		\toprule
		Axis & Dataset \# & 2T (ms) & $\Delta T_a (\mu s)$ & NPoints & $\Phi_0$ (rad) & $\Theta_N$ (rad) & B (rad) & $\chi_{\mathrm{red}}^2$ \\
		\midrule
		X & 0  & 800  & 60 & 41 & 221.572(9)  & 0.908341(41) & 0.787(6)  & 3.9\\
		  & 2  & 800  & 60 & 24 & 221.572(16) & 0.908015(73) & 1.004(12) & 5.0\\
		  & 4  & 801  & 60 & 24 & 222.409(12) & 0.908125(57) & 0.979(8)  & 3.3\\
		  & 6  & 800  & 40 & 42 & 221.569(16) & 0.907845(72) & 0.731(11) & 11.6\\
		  & 8  & 800  & 60 & 46 & 221.524(9)  & 0.908029(43) & 0.833(7)  & 2.6\\
		  & 10 & 800  & 60 & 41 & 221.587(7)  & 0.909072(40) & 0.972(6)  & 2.7\\
		  \midrule
		Y & 1  & 800  & 60 & 41 & 221.545(9)  & -0.661596(39) & 0.628(6) & 3.2\\
		  & 3  & 800  & 60 & 24 & 221.548(23) & -0.662011(95) & 0.678(15)& 14.3\\
		  & 5  & 801  & 60 & 24 & 222.389(20) & -0.662023(83) & 0.665(14)& 12.3\\
		  & 7  & 800  & 40 & 41 & 221.584(16) & -0.662582(74) & 0.690(11)& 13.2\\
		  & 9  & 800  & 60 & 43 & 221.557(11) & -0.662044(51) & 0.545(8) & 6.3\\
		  & 12 & 800  & 60 & 41 & 221.559(7)  & -0.660894(31) & 1.238(5) & 2.5\\
		\bottomrule
	\end{tabular}
	\caption{\textbf{Detail of the 6 datasets taken for the two axis with the fitted parameters and their extracted uncertainties.} Are presented the parameters for each measurements including the total interrogation time 2T, the time asymetry used to prevent the parasitic interferometers from recombining $\Delta T_a$, the number of points per datasets N, the fitted values of the control parameters $\Phi_0$, $\Theta_N$, and B and the corresponding reduced chi-squared for the least-square fit.}
	\label{table:tableDatasets}
	\end{table}
	
In the table \ref{table:tableDatasets} are presented the details and the results for the fitted parameters for all 12 measurements. The parameters that have been varied between datasets are T, $\Delta T_a$ and the number of points per sets. The fitted parameters corresponds to $\Phi_0$ (the amplitude of the sine), $\Theta_N$ the angle where the interferometer is pointing North, and the offset B. Next to this is presented the reduced chi-squared, as an indicator of the quality of the fit.

The angle between the two axes of the gyroscope is found stable at $89.9599(94)^{\circ}$, which represents a deviation of 0.7 mrad with respect to $90^{\circ}$ (consistent with mechanical tolerance).

\section*{Section S2: Scale factor estimation}

\subsection*{Measurement of $T$}
Measurement of the duration of the interferometer has been done with a high resolution oscilloscope (Textronix 4 series Mixed signal oscilloscope) measuring directly the time between the Raman pulses photodiodes signals and yields a result of 400.0020(1) ms.

\subsection*{Measurement of $g$}
Measurement of g has been performed at the position of the experiment with a cold atom gravimeter (CAG). Value has been corrected from the vertical gradient, accounting for an experiment at 1 m height.

\subsection*{Measurement of $\theta_0$}
To lift the degeneracy between $+\keff$ and $-\keff$, the Raman beams are tilted by an angle $\theta_0$.
The accurate measurement of this angle is necessary to estimate the expected scale factor value of the gyroscope.

One way to estimate the angle is to keep the 4 pulses geometry and change both the time asymmetry $\Delta T_a$ and the rate of the Doppler compensating ramp $\alpha$.
The ramp is necessary to maintain the resonance condition while the atoms are moving, as the Doppler detuning rise linearly. 
Expected phase shift is 

\begin{equation}
	\Phi = 2T \Delta T _a\left(k_{\mathrm{eff}} g \sin(\theta_0) - \alpha \right)
\end{equation}
By finding the ramp that cancel the phase shift, one can extract the angle of the Raman beams, in the same way that compensating the ramp is used to determine $g$ on the cold atom gravimeters.

\begin{figure}[htbp]
	\centering
	\includegraphics[width=0.9\columnwidth]{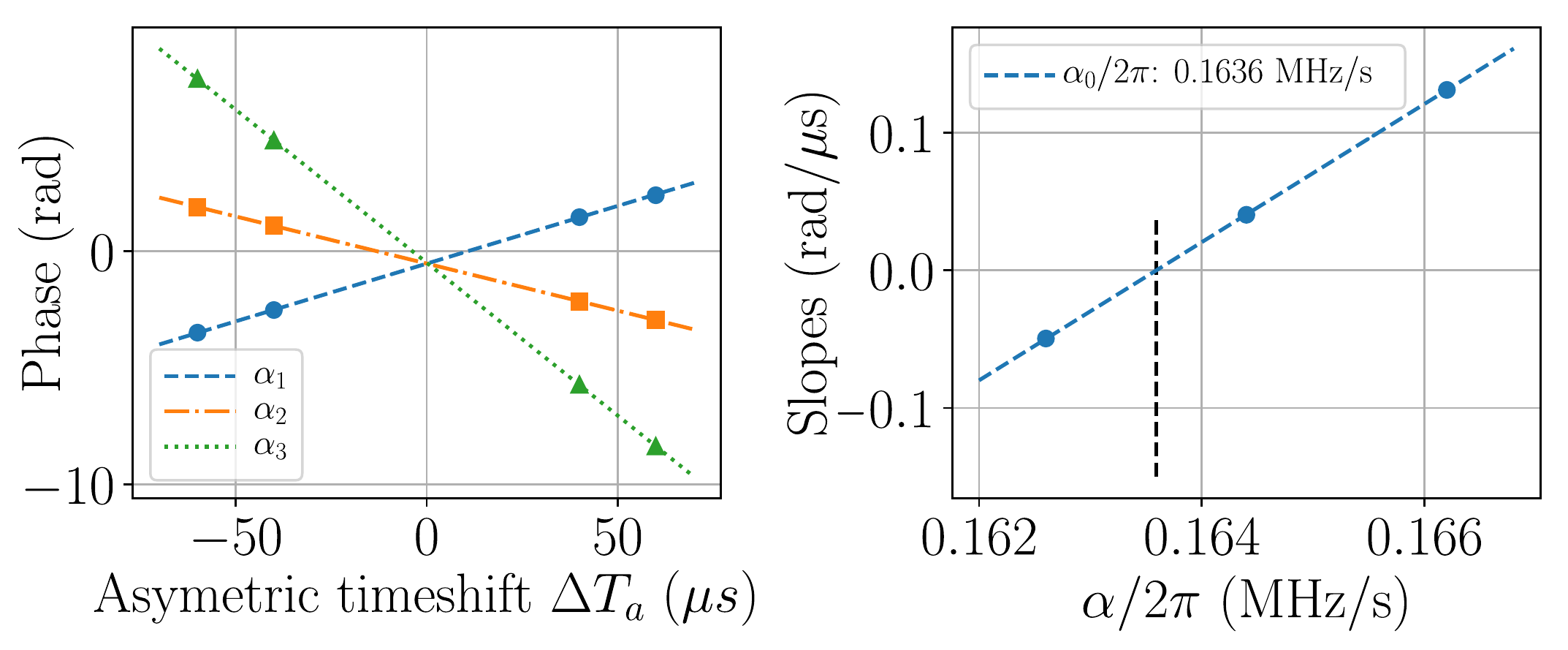}
	\caption{\textbf{Measurements of the angle of the Raman collimators.} On the left is presented the phase shifts versus the asymmetric time shift for 3 different ramp rate. On the right is the slopes of the previous curves versus the ramp rate applied. The Ramp where the slope is 0 is the one perfectly compensating the Doppler effect and is used to calculate the angle.}
\end{figure}

The measurements yields the following results:  $\theta_0^x = 4.0750(5)^{\circ}$ , and $\theta_0^y = 4.1251(3)^{\circ}$.

\subsection*{Measurement of $\deltaK$}

\begin{figure}[htbp]
\centering
\includegraphics[width=0.7\columnwidth]{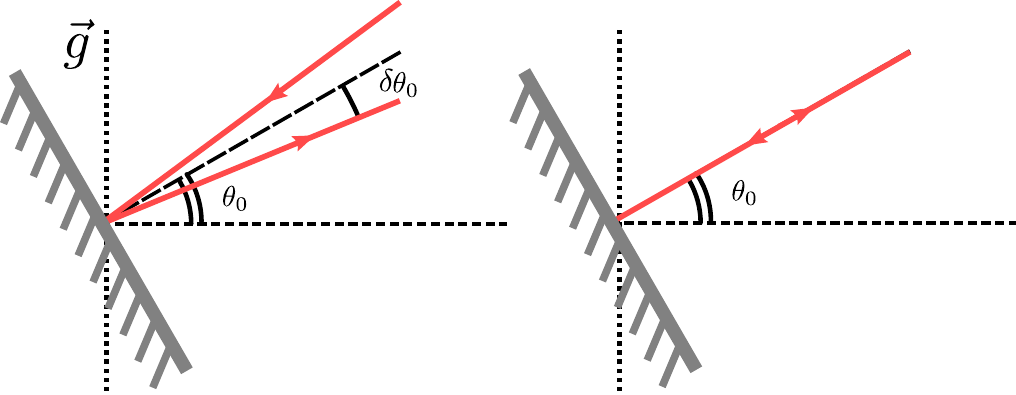}
\caption{\textbf{Illustration of the effect of a misalignment of one beam on the modulus of $k_{\mathrm{eff}}$.} Side by side are presented a misaligned wave vector and a perfectly aligned one.}
\label{fig:nek}
\end{figure}

As studied in (32), a different angle between the two Raman collimators lead to a different modulus of the $\keff$ vector of both Raman beams, as illustrated on figure \ref{fig:nek}.

This $\deltaK$ between the two beams is defined as $k_{\mathrm{eff}}^{(B)} - k_{\mathrm{eff}}^{(T)}$, where $\keff ^{(B)}$ represents the wave vector of the bottom beam and $k_{\mathrm{eff}}^{(T)}$ the one from the top beam.
This $\deltaK$ introduce a correction to the scale factor of the experiment and prevents the interferometer from closing correctly.

In presence of a $\deltaK$ and for a 4 pulses interferometer, the two middle pulses needs to be shifted symmetrically by $\Delta T_s$ with respect to the apogee to close the interferometer. If the closing condition:

\begin{equation}
	\Delta T_s \simeq \frac{T \deltaK}{ 2 \keff}
\end{equation}
is not met, the maximum of contrast of the interferometer will be shifted in time.
In a way to deduce the difference of $\keff$ between the two beams, we measure the contrast while varying the symmetric timeshift $\Delta T_s$.

\begin{figure}[htbp]
	\centering
   		\includegraphics[width=0.8\columnwidth]{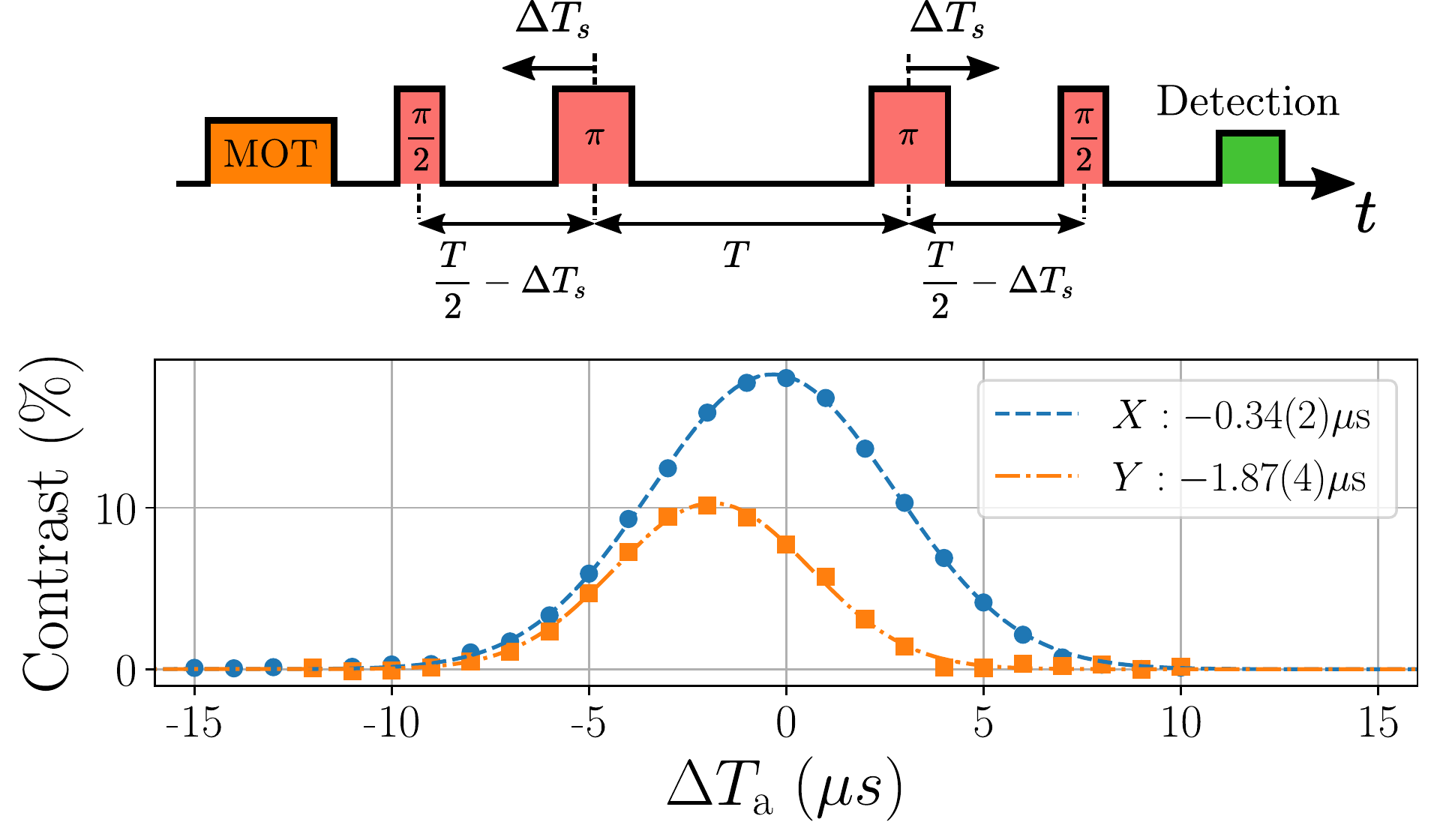}
	\caption{\textbf{Illustration of the sequence used to measure the $\deltaK$ with the 4 pulses interferometer.}
	Contrast for X and Y axis (respectively blue dots and orange squared points) versus the displacement of the two central pulses by $\Delta T_s$. Blue dashed (orange point-dashed) lines  are Gaussian fits to the data.}
\end{figure}
For the X axis, the measured shift is  $-0.34(2) \mu \mathrm{s}$ and for the Y axis , the measured timeshift is $-1.87(4) \mu \mathrm{s}$, corresponding respectively to a $\deltaK$ of $-25(1) m^{-1}$ for X and $-138(3) m^{-1}$ for Y. 
The origin behind the non negligible shift in the Y axis is of a technical nature, the Raman collimators for the X axis are outside the magnetic shield and the retro-reflection can be aligned by placing a target in front of it. The study of this effect and it's influence had been done on this easy to adjust axis (32).
The Y collimators on the other hand are inside the two layers of magnetic shield and the misalignment cannot be easily corrected.

Presence of a $\deltaK$ will introduce a correction to the scale factor of the gyroscope associated with an imperfect recombination of the arms of interferometer. It is possible to reduce the corrections by adjusting the timings of the central $\pi$ pulses with a $\Delta T_s$ to be at the maximum of contrast, in the same way that we measured it previously with the 4 pulses interferometer.

Once the timings are adjusted and taking into account the uncertainty on the value of $\epsilon$, the full scale factor becomes:
\begin{equation}
\Phi_{\Omega}(\Theta)=\frac{T^3}{2} \keff (1 - \frac{2}{3}\epsilon - \frac{11}{3}\delta \epsilon) g \cos(\theta_0) \times  \cos(\psi) \Omega_E \times \cos(\Theta - \Theta_N)
\label{eq:sagnac_expandedDeltaK}
\end{equation}

where $\epsilon = \deltaK/\keff$ is the measured value for the misalignment of the two beams, and $\delta \epsilon$ the uncertainty on this measurement.

For the X axis ,  $\epsilon$ is found to be $-1.7(1)\times 10^{-6}$ and $-9.3(2) \times 10^{-6}$ for Y. Respectively a relative contribution to the scale factor of $7.7 \times 10^{-9}$(0.07 ppm) for X and $4.2\times 10^{-8}$(0.13 ppm) for Y.

\subsection*{Measurement of $\Psi$}
We used GNSS data to find the geographic latitude our experiment is located at. This latitude is defined as perpendicular to the ellipsoid of reference which include both Earth's mass density model distribution and centrifugal acceleration parts. This geodetic latitude needs to be corrected to take into account local Earth inhomogeneities and altitude. Two corrections appears, the North deviation and the East deviation. However, we are only interested in the North deviation as it will change the projection of the Earth rotation vector on our experiment.

\begin{figure}[htb]
	\centering
	\includegraphics[width=0.4\columnwidth]{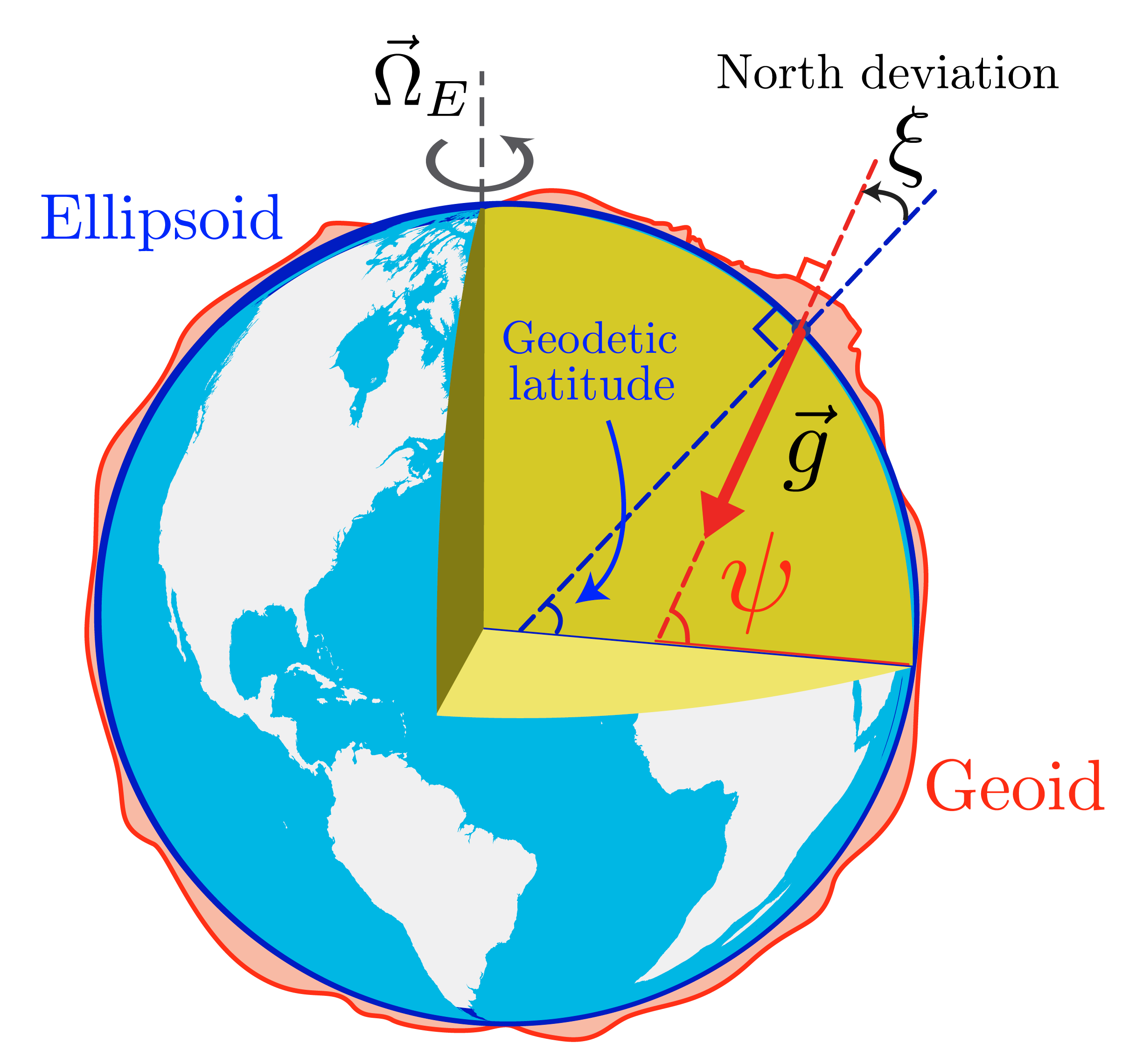}
	\caption{\textbf{Astronomical latitude.} Illustration of the Earth’s geoid (in red) and ellipsoid (in blue). The geoid is the equipotential surface of the Earth’s gravity field and the ellipsoid is the approximate shape of the Earth.}
\end{figure}

Calculations of the deviations have been done by the IGN Service de G\'eodésie et de M\'etrologie, and make use of the QGF16 quasi-geoid grid.
\section*{Section S3: Phase shift calculations and second order terms}

The full calculation of the inertial phase shift yields other terms than the main Sagnac phase shift, mainly second order terms that are negligible at our level of sensitivity, but that would need to be considered if we were targeting ppm level measurements.
The trajectories of the atoms have been calculated by solving the equation of motion taking gravity, Coriolis and centrifugal forces into account.
The calculated positions and velocities are used to calculate the total phase shift at the output of the interferometer, with the timings and geometry of the experiment.
From the equation of motion is derived the position of the center of mass of the atoms at time t in the laboratory frame: $x(t)$, $y(t)$, and $z(t)$.
From this the phase shift becomes:
\begin{equation}
	\begin{aligned}
	\Phi &= \phi_1 - 2 \phi_2 + 2\phi_3 - \phi_4 \\
		 &= \vec{k}_{\mathrm{B}} . \left[ \vec{r}\left(0\right) 
		 -  \vec{r}\left(2T\right) \right] + 		\vec{k}_{\mathrm{T}} . \left[ - 2 \vec{r}\left(\frac{T}{2}\right) 
		 + 2 \vec{r}\left(\frac{3T}{2}\right)\right]
	\end{aligned}
\end{equation}
where the two effective wave vectors for the top beam ($k_\mathrm{T}$) and the bottom beam ($k_\mathrm{B}$) are distinguished as to take into account the possible effects on the trajectory if they are found not to be equal.

The phase shift has been developed in order of $\Omega$ and the terms are presented in table \ref{table:phaseShifts}, with their relative values compared to the main Sagnac phase shift (In the table with a contribution of 1). 

Secondary terms include:\textit{(i)} coupling between inaccurate trajectory ($v_{x0}$, $v_{y0}$ and $v_{z0}$) and the residual $\keff$ misalignment represented by $\epsilon$; \textit{(ii)} coupling between the introduced time asymmetry $\Delta T_a$ and $\epsilon$; \textit{(iii)} Second order terms in $\Delta T_a$,  \textit{(iv)} second order terms in $\Omega^2$ coupled with $\Delta T_a$ and finally \textit{(v)} recoil terms that appear by taking the recoil of the atoms into account for their trajectories. Note that in the perfect case of no time asymmetry $\Delta T_a$ and a zero misalignment between the wave-vectors, only the Sagnac phase shift and the recoil terms will remain.

One of the second order term is the one due to $\Delta T_a$ which is not compensated by alternating $\pm \Delta T_a$ measurements:

\begin{equation*}
	2T \Delta T_a^2 k_{\mathrm{eff}} g \Omega \cos(\psi) \cos(\theta_0) \sin(\theta_N)
\end{equation*}
This term has a relative contribution of $9 \times 10^{-8}$ for a typical $\Delta T_a$ of $60 \mu s$.

Contributions from the finite pulse duration needs to be taken into account at the level of few $ 10^{-9}$ or below, and are not presented here.

Fluctuation of the modulus of g due to the tides can modify the scale factor at the level of $3\times  10^{-7}$, but can be taken into account with an accurate tide model like it is done for gravity measurements.

Terms due to vertical and/or horizontal gravity gradient can appear, but at the level of few $10^{-7}$.

Fluctuations of the position of the geographical North, principally the Chandler and annula wobble have been estimated in (48) and account for 20 to 100 of $\text{p-rad}.s^{-1}$, below the current accuracy of our instrument.

Below is presented a table recapitulating the second order terms and their respective contribution to the total phase shift.
All of those term can be safely neglected at the level of accuracy of our experiment.

\begin{table}[htbp]
\begin{center}
\begin{tabular}{l|S[table-format=2.2e-2]}
Terms independent of $\Theta$ (Bias terms) & {Absolute Phase value (rad)}\\
\hline
\hline
\rowcolor{gray!20}
 $-2 \: \keff\:g\: T \sin(\theta_0)\: \Delta T_a$ & -5.1e2\\
\rowcolor{gray!20}
$+2 \: \keff\:g\: T \sin(\theta_0)\: \Delta T_a\: \epsilon$ &   4.9e-5\\ 
&\\
$-2 \:\keff \:T\: \cos(\theta_0)\: \epsilon\: v_{x_{0}} $ &  -1.2 e-1 \\
$+2 \:\keff \:T\: \sin(\theta_0)\: \epsilon\: v_{z_{0}} $ &  8.4e-9 \\

$\rule{0pt}{2ex}-4\: \keff \: T^{2}\: \Omega\:  \cos(\theta_0)\: \sin(\psi)\:\epsilon\: v_{y_{0}}  $ & -5.1e-6  \\
\rowcolor{gray!20}
$-4\: \keff\: T\: \Omega\:   \cos(\theta_0)\: \sin(\psi)\: \Delta T_a\: v_{y_{0}} $ & -7.7e-6  \\
& \\
$-\frac{\hbar}{m}\: \keff^2\: T^3 \:\Omega^2\: \cos(\theta_0)^2 \: \cos(\psi)^2$  & -1.5e-5\\
$+\frac{\hbar}{m}\: \keff^2\: T^3 \:\Omega^2\: (\cos(\theta_0)^2 + \cos(\psi)^2)$  & 5.1e-5\\
&\\

\end{tabular}
\end{center}
\caption{\textbf{Bias terms coming from the full calculation of the Sagnac phase shift and their values.}
Terms highlighted in gray rows are terms which will be canceled while alternating the sign of $\Delta T_a$.}
\label{table:phaseShiftsBias}
\end{table}

\begin{table}[htbp]
\begin{center}
\begin{tabular}{l|S[table-format=2.2e-2]}
Terms proportional to $\Omega$ & {Relative phase value}\\
\hline
\hline
$-\frac{1}{2}\:   \keff\:g\: T^{3}\: \Omega\: \cos(\theta_0) \:\cos(\psi) \:\sin(\Theta) $  & 1  \\
&\\
$+\frac{11}{6}\:  \keff\:g\: T^{3}\: \Omega\: \cos(\theta_0) \:\cos(\psi) \:\sin(\Theta) \:\epsilon$  & 3.7e-5  \\
& \\
\rowcolor{gray!20}
$+4\: \keff\: T\: \Omega\: \cos(\theta_0)\: \cos(\psi)\: \sin(\Theta)\: \Delta T_a\: v_{z_{0}} $ &  3.1e-7  \\
\rowcolor{gray!20}
$+4\: \keff\: T\: \Omega\:   \sin(\theta_0)\: \cos(\psi)\: \cos(\Theta)\:\Delta T_a\: v_{y_{0}}  $ &  2.2 e-8 \\

& \\
$+2\:  \keff\: T\: \Omega\:g\:  \cos(\theta_0) \:\cos(\psi) \:\sin(\Theta)  \:\Delta T_a^{2} $  & 9.0e-8 \\
&\\
$+4\: \keff\: T^{2}\: \Omega\:  \cos(\theta_0)\: \cos(\psi)\: \sin(\Theta) \:\epsilon \: v_{z_{0}}  $  & 2.0e-7  \\

$+4\: \keff\: T^{2}\: \Omega\:  \sin(\theta_0)\: \cos(\psi)\: \sin(\Theta)\:\epsilon \: v_{x_{0}}  $  & 1.5e-9  \\

& \\
Terms proportional to $\Omega^{2}$ & \\
\hline
\hline
\rowcolor{gray!20}
$-\frac{11}{3}\: g\: \keff\: T^{3}\: \Omega^{2}\: \cos^{2}(\psi)\: \sin(\theta_0)\: \Delta T_a $  & -1.5e-9  \\
\rowcolor{gray!20}
$+\frac{11}{3}\: g\: \keff\: T^{3}\: \Omega^{2}\: \cos(\psi)\: \sin(\psi)\: \cos(\theta_0)\: \cos(\Theta)\: \Delta T_a $  & 2.2 e-8  \\
&\\
$-\frac{2\hbar}{m}\: \keff^2\: T^3 \:\Omega^2\: \sin(\theta_0) \: \cos(\theta_0) \: \cos(\psi) \:\cos(\Theta)$  & -1.1e-8\\
$-\frac{\hbar}{m}\: \keff^2\: T^3 \:\Omega^2\: \cos(\theta_0)^2 \: \cos(\psi)^2 \: \cos(\Theta)^2$  & -6.8e-8\\
\end{tabular}
\end{center}
\caption{\textbf{First and second order terms coming from the full calculation of the Sagnac phase shift and their relative value compared to the first order approximation that is being used in the main text.} The terms appear to be coupling between the different asymmetries of the experiment, for example coupling between velocity errors ($v_{x0}$, $v_{y0}$ and $v_{z0}$) and the asymmetric timing $\Delta T_a$. We find as well couplings between $\epsilon$ due to the misalignment of the Raman beams and $\Delta T_a$ or velocity errors. If the full trajectories of the atoms are taken into account, appears some recoils terms both as bias phases and terms proportional to $\Omega ^2$. Higher order terms appears but only up to the second orders are presented here for clarity. Terms highlighted in gray rows are terms which will be canceled while alternating the sign of $\Delta T_a$.}
\label{table:phaseShifts}
\end{table}

\section*{Section S4: Drift of the bias - Statistical simulation}

During week long measurements, we can observe fluctuations of the alignment of the mirrors and of the atomic trajectories, creating a bias term drifting in time:

\begin{equation}
\begin{aligned}
\Phi(t) &= 2T\keff \Delta\theta(t) \Delta v(t)\\
&\approx 12 \mathrm{mrad/\mu rad/mm.s^{-1}}
\end{aligned}
\end{equation}
where $\Delta \theta (t)$ represents represents the mirror misalignment and $\Delta v (t)$ the error on the trajectory.

A visible consequence of mirror misalignment is a change on the contrast of the interferometer.
Before any measurement point, we measure the contrast of the interferometer to have some insight on it's drift between measurements.

\begin{figure}[tbp]
\centering
\includegraphics[width=0.9\columnwidth]{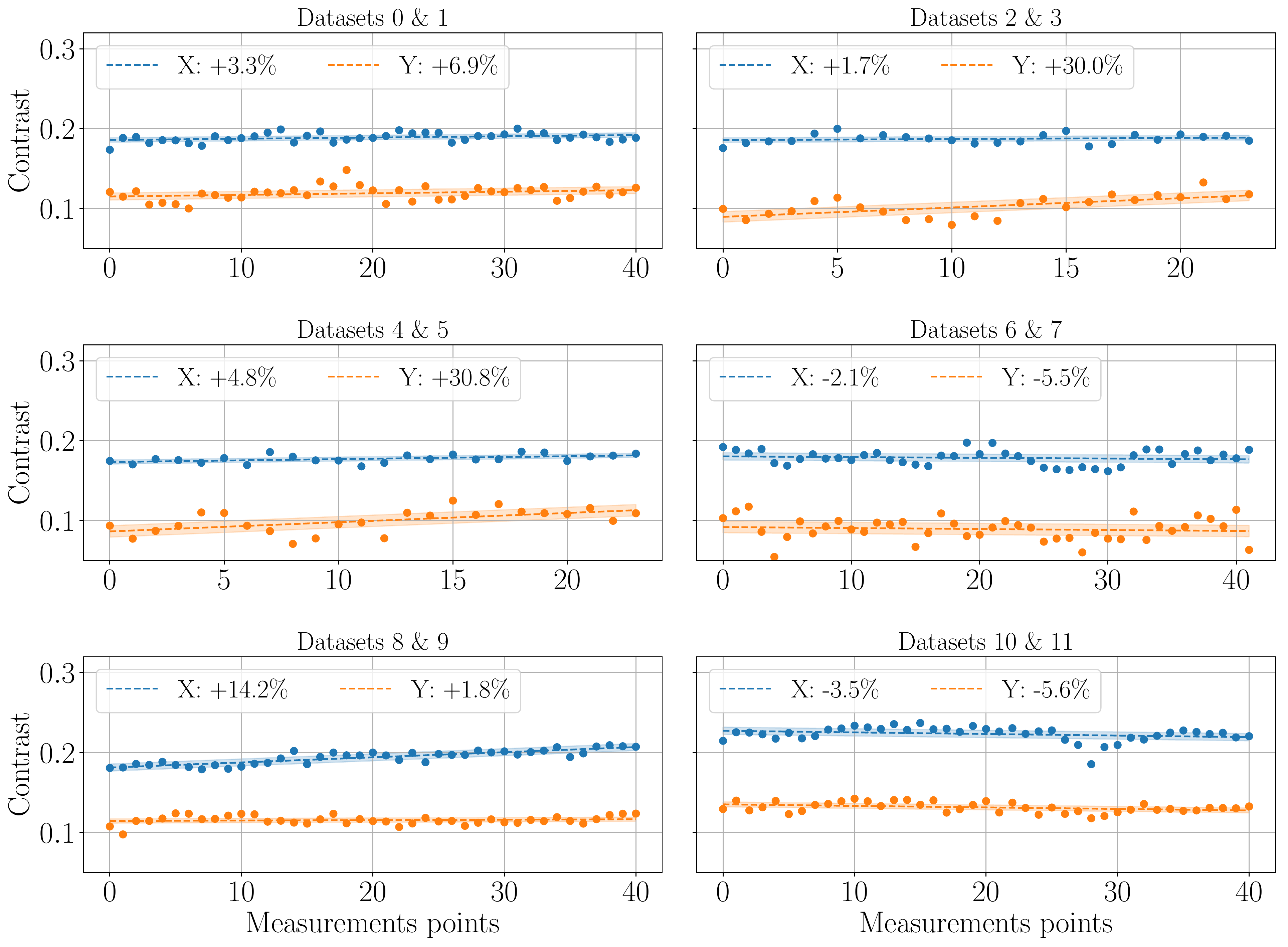}
\caption{\textbf{Measures of the contrast for all datasets.} Each subplots presents the contrast for the X axis (Blue points, dashed blue line) and the Y axis (orange points, dashed orange lines), with a linear adjustment to estimate the drift during the measurement}
\label{fig:contrast}
\end{figure}

When we look at the contrast evolution for each datasets (See Fig \ref{fig:contrast}) we can see long variations on the order of $5\%$ to $30\%$ , probably linked to temperature fluctuations of the laboratory room impacting the mirror relative alignment.
From this level of fluctuation we can deduce that the misalignment has to reach from 3 $\mu$rad to 9 $\mu$rad, knowing the evolution of the contrast with the misalignment (31).

On the other side, typical fluctuations of the launch velocity are of the order of $\sigma_v$ = 1.0 mm.s$^{-1}$ after alignment of the trajectories.

\begin{figure}[htbp]
	\centering
	\includegraphics[width=1\columnwidth]{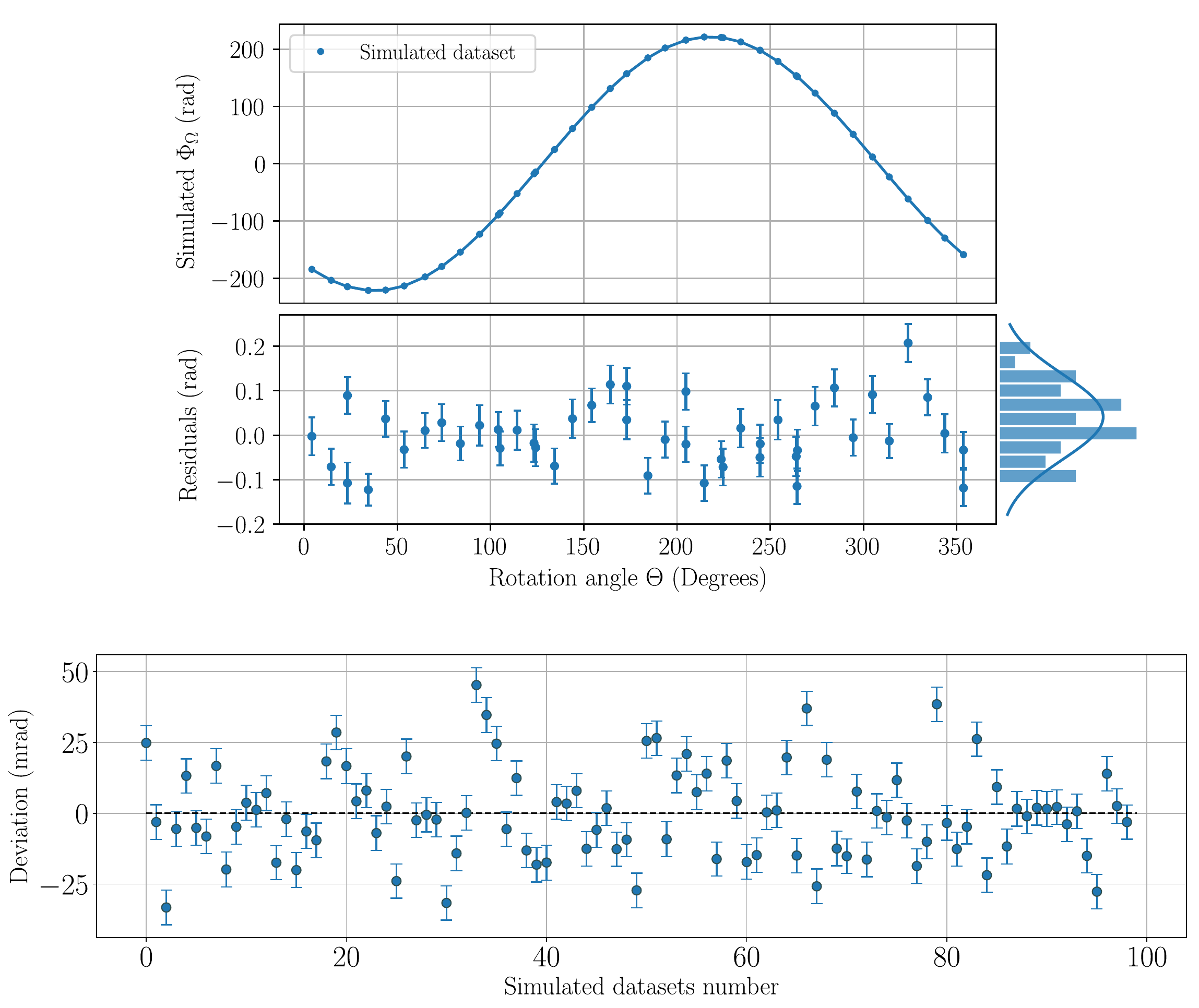}
	
	\caption{\textbf{Example of a simulated dataset and results of the simulation for 99 realizations.}
Third subset represents deviation from the expected Sagnac phase for 99 simulated datasets, showing a dispersion bigger than the statistical noise, and with the order of magnitude matching the real case.}
\label{fig:simulation}
\end{figure}

We use this estimation of the misalignment of both the mirror and the trajectories to simulate the impact of this effect on the measure of the Sagnac phase, which is on the order of 50 mrad. Datas are generated for experiment angles from 0 to 360 $^{\circ}$ from a sine of amplitude equal to the expected Sagnac phase. The fluctuations of the bias are generated from the typical variations and then applied randomly as bias fluctuations. Results for the simulations are presented on figure \ref{fig:simulation} and are to be compared with the measured dataset.

The simulation show that a fluctuation of the bias during the measurements (due to the phase shift linked to the alignment of the mirrors and trajectory) can create a dispersion of the data bigger than the statistical uncertainty. Some simulated datasets can reach up to 50 mrad deviation with the parameters presented before, which could explain the deviation of some of our measured values (mostly the dataset 8, of the 31 May 2021, which is close to 50 mrad away from the expected value).

The level of misalignment and fluctuations of the trajectory needed to reproduce the effect is on the level that we can expect as well, showing that our observed dispersion can be explain by this effect.
	
\end{document}